\documentstyle[aps,12pt]{revtex}
\begin{document}
\title{{\large {\bf Spin and Flavor Contents of the Proton}\thanks{%
\ Presented by Ling-Fong Li at the {\em Ames Symposium }, May 22-24, 1995.}}}
\author{{\normalsize Ling-Fong Li$^1$ and T. P. Cheng$^2$}}
\address{\tighten $^1${\em Department of Physics, Carnegie-Mellon University,
Pittsburgh, PA
15213}\\
$^2${\em Department of Physics and Astronomy, University of Missouri, St
Louis, MO 63121}\\
Report No: CMU-HEP 95-08, DOE-ER/40682-98, hep-ph/9506397}
\maketitle

\begin{abstract}
\begin{center}
{\bf Abstract}
\end{center}

\tighten After a brief review of the experimental results obtained in deep
inelastic lepton-nucleon scatterings and the Drell-Yan processes and their
implications for the spin and flavor contents of the nucleon, we suggest
that those features, that are contrary to the expectations of the naive
quark model, can all be accounted for in the chiral quark model of Georgi
and Manohar. In our formulation of the chiral quark model, the $\eta
^{\prime }$ meson is seen to play an important role.\smallskip\
\end{abstract}

\medskip\ \ \tighten
A few years back, the deep inelastic scattering experiments performed by the
EMC\cite{emc} and NMC\cite{nmc} collaborations at CERN suggested a proton
spin and flavor structure that was at variance from the naive quark model
expectations. These results have been confirmed and extended by more recent
experimental findings: by SMC at CERN, by E142 and E143 at SLAC\cite{nuspin}%
, and by NA51 at CERN\cite{na51}, respectively. In this paper, we shall
discuss a mechanism\cite{ehq}\cite{clprl}, in the framework of the chiral
quark model of nonperturbative QCD\cite{mgtheor}, which gives a simple and
unified account of all such ''anomalous'' spin and flavor structures of the
proton.\\

{\bf 1. Structure Function }$g_1(x)${\bf \ and the Proton Spin}\\

{\em 1.1\ \ Polarized\ lepton-nucleon scatterings and their QCD analysis}

Inclusive deep inelastic{\em \ }$lN$ scatterings basically measure the
current-current correlation
\begin{equation}
{\sl Im}\int e^{iq\xi }d^4\xi \left\langle N({\bf p},s)\left| T\left( J_\mu
^{em}\left( \xi \right) J_\nu ^{em}\left( 0\right) \right) \right| N({\bf p}%
,s)\right\rangle =\frac i{M\nu }\varepsilon _{\mu \nu \alpha \beta }q^\alpha
s^\beta g_1^N(x,Q^2)+...
\end{equation}
where $s_\mu $ is the covariant spin vector of the nucleon, $Q^2\equiv -q^2$
is momentum transfer, and $x=Q^2/2M\nu $ is the usual scaling variable. The
spin-dependent structure function $g_1^N(x,Q^2)$ of the nucleon as defined
above can be measured by comparing the cross sections of the parallel and
anti-parallel longitudinally polarized scatterings. In the operator product
expansion of the left-hand-side, one of the leading twist-two terms with the
correct quantum number is the dimension-three axial-vector current operator $%
A_\mu ^{}(0)=\sum_{q=u,d,s}e_q^2\overline{q}\gamma _\mu ^{}\gamma _5q$. This
term can be isolated by taking the appropriate moment of the structure
function. For the case of the proton target, this leads to the sum rule:
\begin{equation}
\int_0^1dxg_1^p(x,Q^2)=\frac{C_{NS}(Q^2)}{36}\left( 3g_A+\Delta _8\right) +%
\frac{C_S(Q^2)}9\Delta \Sigma  \label{g1sumrule}
\end{equation}
The factors $g_{A,\;}\Delta _{8,}\;$and $\Delta \Sigma $ are the various
flavor non-singlet and singlet combinations of $\Delta q\,s$ which are the
axial vector current matrix elements between the proton states:
\begin{equation}
\left\langle p({\bf p},s)\left| \overline{q}\gamma _\mu ^{}\gamma _5q\right|
p({\bf p},s)\right\rangle \equiv 2s_\mu ^{}\Delta q.  \label{delqdef}
\end{equation}
$\Delta \Sigma =$ $\Delta u+\Delta d+\Delta s$ is the singlet combination,
while the non-singlet combinations can be related, via $SU(3),$ to the axial
vector couplings measured in the baryon weak decays:
\begin{eqnarray}
\Delta u-\Delta d &=&g_A=1.2573\pm 0.0028  \nonumber  \label{su3values} \\
\Delta u+\Delta d-2\Delta s &=&\Delta _8=3F-D=0.61\pm 0.038.
\label{su3values}
\end{eqnarray}
The Wilson coefficients in Eq.$\left( \ref{g1sumrule}\right) $ have been
calculated in perturbative QCD\cite{qcdc}, for the non-singlet case, up to
three loops, with an estimate made for the fourth order term:
\begin{equation}
C_{NS}(Q^2)=1-\frac{\alpha _s(Q^2)}\pi -3.5833\left( \frac{\alpha _s(Q^2)}\pi
\right) ^2-20.2153\left[ \frac{\alpha _s(Q^2)}\pi \right] ^3+O(130)\left[
\frac{\alpha _s(Q^2)}\pi \right] ^4,
\end{equation}
for the singlet-channel, to the second, and estimated to the third, order:
\begin{equation}
C_S(Q^2)=1-\frac{\alpha _s(Q^2)}\pi -1.0959\left( \frac{\alpha _s(Q^2)}\pi
\right) ^2-O(6)\left[ \frac{\alpha _s(Q^2)}\pi \right] ^3.
\end{equation}
The $Q^2$-dependence appearing through the $\alpha _s(Q^2)$ is important to
reconcile the various experimental results taken at different $Q^2s$.\cite
{ellisk}\footnote{%
These authors also conclude that the Bjorken sum rule $\int dx(g_1^p-g_1^n)=%
\frac 16g_A$ is satisfied to within 12\%.}

Thus the $g_1(x)$ measurements can be used to evaluate the moment-integral.
This, together with the $SU(3)$ values as given in Eq. (\ref{su3values}),
allows us to deduce the axial vector matrix element for each quark flavor $%
\Delta q$. In the following we display both the original and the more recent
results:

\begin{center}
\begin{equation}
\begin{tabular}{cccc}
$Experiments$ & $\;\int_0^1dxg_1^p(x,Q^2)$ & $\Delta s$ & $\Delta \Sigma $
\\ \hline\hline
&  &  &  \\
$EMC\;(1987)$ & $\;0.126\pm 0.025$ & $\;-0.19\pm 0.06$ & $\;\;0.12\pm 0.17$
\\
&  &  &  \\
$SMC,\;E142$-$3\;(1994)$ & $\;0.136\pm 0.016$ & $\;-0.10\pm 0.04$ & $%
\;\;0.31\pm 0.11$%
\end{tabular}
\label{smc}
\end{equation}
\end{center}

We can thus conclude that the new data do support the original findings of a
significant strange quark contribution $\Delta s\neq 0$\cite{ejsr} and a
much-less-than-unity total-quark term $\Delta \Sigma <1,$ although the size
of $\Delta s$ has decreased somewhat and $\Delta \Sigma \,$ is perhaps not
as small as originally thought.\ \\

{\em 1.2\ Theoretical interpretation - proton spin contents}

{}From above we see that the $g_1$ - sum rule measures the matrix element of
the axial vector current which corresponds to the non-relativistic spin
operator
\begin{equation}
\overrightarrow{{\bf A}}{\bf =}\overline{q}\overrightarrow{{\bf \gamma }}%
\gamma _5q=q^{\dagger }\left(
\begin{array}{cc}
\overrightarrow{{\bf \sigma }} & 0 \\
0 & \overrightarrow{\sigma }
\end{array}
\right) q.
\end{equation}
Thus we can interpret the $\Delta q$ defined in (\ref{delqdef}) as the quark
contribution to the proton spin.

The phenomenological results of $\Delta s\neq 0$ and $\Delta \Sigma <1$ is
contrary to the simple quark model expectation. The $SU(6)$ wave function
appropriate for the usual constituent quark model of the proton has the form
of
\begin{equation}
\left| p_{\uparrow }\right\rangle =\frac 1{\sqrt{6}}\left( 2\left|
U_{\uparrow }U_{\uparrow }D_{\downarrow }\right\rangle -\left| U_{\uparrow
}U_{\downarrow }D_{\uparrow }\right\rangle -\left| U_{\downarrow
}U_{\uparrow }D_{\uparrow }\right\rangle \right)
\end{equation}
which yields the quark spin contributions of
\begin{equation}
\Delta U=2\times \frac 46=\frac 43,\;\;\;\Delta D=\frac 16\left[ 4\left(
-1\right) +1+1\right] =-\frac 13,\;\;\;\Delta S=0,\;\;\;\Delta \Sigma =1.
\end{equation}

However, we have to distinguish between {\em current quarks} which appears
in the current operators being probed by lepton-proton scatterings and the
{\em constituent quarks} which appear in the SU(6) calculation. Presumably,
constituent quarks are the current quarks dressed up by some nonperturbative
QCD interactions. A simple proposition\cite{kapmano} will be that they are
proportional to each other$\;\left\langle \left| \overline{Q}\gamma _\mu
\gamma _5Q\right| \right\rangle =Z_q\left\langle \left| \overline{q}\gamma
_\mu \gamma _5q\right| \right\rangle $ with these ''renormalization
factors'' $Z_q^{\prime }s\;$still obey the flavor $SU(3)$ symmetry.
\begin{eqnarray*}
\left\langle \left| \overline{U}U+\overline{D}D-2\overline{S}S\right|
\right\rangle &=&Z_8\left\langle \left| \overline{u}u+\overline{d}d-2%
\overline{s}s\right| \right\rangle \\
\left\langle \left| \overline{U}U+\overline{D}D+\overline{S}S\right|
\right\rangle &=&Z_0\left\langle \left| \overline{u}u+\overline{d}d+%
\overline{s}s\right| \right\rangle
\end{eqnarray*}
Thus, even there is no constituent strange quark in the proton\ $%
\left\langle \left| \overline{S}S\right| \right\rangle =0$, the proton
matrix elements of the strange quark bilinears is still nonzero if the
singlet and octect operators renormalize differently
\begin{equation}
\left\langle \left| \overline{s}s\right| \right\rangle =\frac 13\left( \frac
1{Z_0}-\frac 1{Z_8}\right) \left\langle \left| \overline{U}U+\overline{D}%
D\right| \right\rangle .
\end{equation}
This means that the constituent quarks have some structure of their own; the
renormalization effects that turns the current quark into a constituent
quark surround the constituent quarks with a cloud of quark pairs which may
include strange quarks. In Sec. 3 we shall discuss a specific realization of
this possibility in the framework of the chiral quark model.\\

{\bf 2.\ \ Anomalous Flavor Structure of the Proton}\\

{\em 2.1\ \ Flavor asymmetry measurements in DIS processes}

The difference of the proton and neutron (spin-averaged) structure functions
$F_2^p(x)-F_2^n(x)$ can be expressed in terms of the quark densities,
\begin{equation}
F_2^p(x)-F_2^n(x)=\frac x3\left[ \left( u-d\right) +\left( \overline{u}-%
\overline{d}\right) \right] =\frac x3\left[ 2{\cal I}_s+2\left( \overline{u}-%
\overline{d}\right) \right]
\end{equation}
where we have used isospin symmetry in the first equation, and used, in the
second equation, the definition ${\cal I}_s=\frac 12\left( u-d\right) -\frac
12\left( \overline{u}-\overline{d}\right) ,$ with its integral being the
third component of the isospin: $\int_0^1{\cal I}_sdx=\frac 12.$ The simple
assumption that $\overline{u}=\overline{d}\;$in the quark sea, which is
consistent with it being created by the flavor-independent gluon emission,
then leads to the Gottfried sum rule\cite{gsr}
\begin{equation}
I_G\equiv \int_0^1\frac{dx}x\left[ F_2^p(x)-F_2^n(x)\right] =\frac 13.
\label{gsr}
\end{equation}
Experimentally, NMC found that, with a reasonable extrapolation in the very
small-$x$ region, the integral $I_G$ deviated significantly from one third%
\cite{nmc}.
\begin{equation}
I_G=0.235\pm 0.026=\frac 13+\frac 23\int_0^1\left( \overline{u}-\overline{d}%
\right) dx  \label{nmc}
\end{equation}
This translates into the statement that, in the proton quark-sea, there are
more down-quark pairs as compared to the up-quark pairs.

There were suggestions that the small-$x$ extrapolation could be incorrect
and that a more direct confirmation of $\overline{u}\neq \overline{d}$ at a
fixed $x$ value would be a measurement of the difference of the Drell-Yan
process of proton on a proton vs. neutron targets\cite{es-dy}. NA51
collaboration at CERN has performed such an experiment and found that at $%
x=0.18$ the $\overline{u}$ density is only about half as much as the $%
\overline{d}$ density\cite{na51}:
\begin{equation}
\overline{u}/\overline{d}=0.51\pm 0.04\pm 0.05.  \label{na51}
\end{equation}

\{{\em Remarks:\ }In this discussion of the anomalous flavor content of the
proton, we should also recall the old $\pi N$ sigma-term problem\cite
{tpsigma}. A simple $SU(3)$ calculation, that is entirely similar to that
discussed above in connection with the proton spin, can relate $\sigma _{\pi
N}$ to the ''fraction of strange quarks in the proton'' as:
\begin{equation}
f_s\equiv \frac{2\overline{s}}{3+2\left( \overline{u}+\overline{d}+\overline{%
s}\right) }\simeq \frac{\sigma _{\pi N}-25\,MeV}{3\sigma _{\pi N}-25\,MeV}
\label{sfrac}
\end{equation}
where in the second (approximate) equality we have plugged into the $SU(3)$
result the ''octect-baryon-mass'' value of $M_8=\frac 13(2M_N-M_\Xi
-M_\Sigma )=M_\Lambda -M_\Xi \simeq -200\,MeV$ and the current-quark-mass
ratio of $m_s/m_{u,d}=25.$ Thus the generally accepted value\cite{sigma} of $%
\sigma _{\pi N}=45\,MeV$ translates into a surprisingly large strange quark
fraction of $f_s\simeq 18\%.$\}\\

{\em 2.2\ \ Theoretical interpretation - proton flavor contents}

The simple picture that quark pairs are produced by the flavor-independent
gluons would suggest $\overline{u}=\overline{d}.$ From this viewpoint, the
results of (\ref{nmc}) and (\ref{na51}) are very surprising.

{\em Feynman and Field}\cite{ff} have pointed out long ago that this
equality would not strictly hold even in perturbative QCD, because the Pauli
exclusion principle and the $u$ - $d$ valence-quark asymmetry in the proton
would bring about a suppression of the gluonic production of $\overline{u}\,$%
s (versus $\overline{d}$ s). This mechanism is difficult to implement as the
parton picture is intrinsically incoherent and it is difficult to see how
this can generate such a large asymmetry as experimentally observed.

{\em Pion cloud mechanism}\cite{henley} is another idea to account for the
violation of Gottfried sum rule. The suggestion is that the lepton probe
also scatters off the pion cloud surrounding the target proton, and the
quark composition of the pion cloud is thought to have more $\overline{d}\,s$
than $\overline{u}\,s$. There is supposed to be an excess of $\pi ^{+}$
(hence $\overline{d}$) compared to $\pi ^{-}$, because $p\rightarrow n+\pi
^{+}$, but not a $\pi ^{-}$ if the final states are restricted to the proton
and neutron. (Of course the neutral pions have $\overline{d}=\overline{u}.$)
However, it is difficult to see why the long distance feature of the pion
cloud surrounding the proton should have such a pronounced effect on DIS
processes which should probe the {\em interior} of the proton, and also this
effect should be significantly reduced by emissions such as $p\rightarrow
\Delta ^{++}+\pi ^{-},etc.$

In the following we shall discuss how a modified version of the
Georgi-Manohar {\em chiral quark model} can accommodate these unexpected
proton flavor structures.\\

{\bf 3.\ Proton Structure in the Chiral Quark Model}\\

{\em 3.1\ \ The SU(3) symmetric chiral quark model}

Georgi and Manohar have suggested\cite{mgtheor} that the successes of the
chiral symmetric description of low energy hadron physics naturally
indicates a chiral symmetry breaking scale of $\Lambda _{\chi SB}\simeq
1\;GeV$, significantly higher than the QCD confinement scale of $\Lambda
_{QCD}\simeq (100$-$300)\,MeV.\;$This means that even inside a hadron the
Nambu-Goldstone bosons are relevant degrees of freedom. Namely, while
quarks, gluons and perturbative QCD fully describe strong interactions at
distances $\ll \Lambda _{\chi SB}^{-1},$ for longer distances, but still
inside the nucleon where the nonperturbative QCD effects are expected to
dominate, the physical description may be quite simple when given in terms
of Goldstone-boson modes coupled to quarks. In this regime the chiral
condensates also supply an extra mass to quarks, giving rise to a
constituent quark mass $O(\frac 13M_N)$. While the QCD coupling is expected
to be so strong as to trigger the nonperturbative effects of chiral symmetry
breaking, the remanent effective gluonic coupling in such a quasiparticle
description may well be so small that they can be neglected.

Eichten, Hinchliffe, and Quigg are the first ones to work out the
consequences of this chiral quark model description for the proton flavor
and spin structures\cite{ehq}. It has some similarity to the pion cloud
approach mentioned above, but without its conceptual difficulties. A
straightforward $SU(3)$ calculation yields the averaged antiquark numbers $%
\overline{q}\,s$ in the proton:
\begin{equation}
\overline{u}=\frac 63a,\;\;\overline{d}=\frac 83a,\;\;\overline{s}=\frac{10}3%
a  \label{qbar}
\end{equation}
where the parameter $a$ has the interpretation as the probability of
Goldstone mode emission in $u\rightarrow d+\pi ^{+}$, or its $SU(3)$
equivalences.

In the chiral quark regime, emission of a Goldstone-boson by quarks will
flip their helicities and thus {\em reduce} their contributions to the
proton spin, while the component-quarks of the Goldstone boson will not
contribute because they are necessarily unpolarized. Again a simple
calculation yields
\begin{equation}
\Delta u=\frac 43-\frac{37}9a,\;\;\Delta d=-\frac 13-\frac 29a,\;\;\Delta
s=-a.  \label{delq}
\end{equation}
Thus with $a\simeq 0.22$ as required by the NMC data (\ref{nmc}), the result
shown above in (\ref{delq}) gives a negative-valued total-quark spin
contribution of $\Delta \Sigma =1-16a/3\simeq -0.17,$ which is barely
consistent with the 1987 EMC data (\ref{smc}).\\

{\em 3.2\ \ The chiral quark model with a broken U(3) symmetry}

We have proposed a broken-$U(3)$ version of the chiral quark model with the
inclusion of the $\eta ^{\prime }$ mode\cite{clprl}.

Our motivation is two fold: It is clear that the above description, without
the $\eta ^{\prime }$ mode, of $\overline{u}/\overline{d}=0.75,\;\Delta
s\simeq -0.22,$ and $\Delta \Sigma \simeq -0.17$ as given in Eqs.(\ref{qbar}%
) and (\ref{delq}) is not adequate to account for the more precise nature of
the new data collected in the last two years, as shown in (\ref{na51}) and (%
\ref{smc}), etc. On the theoretical side, we are motivated to include $\eta
^{\prime }$ because, in the leading $1/N_c$ approximation (the planar
diagrams), there are {\em nine }Goldstone bosons with an $U(3)$ symmetry.
However we also know that if we stop at this order, some essential physics
would have been missed: At the planar-diagram level there is no axial
anomaly and $\eta ^{\prime }$ would have been a {\em bona fide} Goldstone
boson; and the unbroken $U(3)$ symmetry would also lead\cite{ehq}\cite{clprl}
to the phenomenologically unsatisfactory feature of a flavor-independent
sea, $\overline{u}=\overline{d}=\overline{s},$ which clearly violates the
results in (\ref{nmc}) and (\ref{na51}).$\,\,$Thus it will be better to
include the $1/N_c$ corrections (non-planar diagrams) which break the $U(3)$
symmetry. The broken-$U(3)$ is implemented by taking
differently-renormalized octet and singlet Yukawa couplings $%
g_0/g_{8.}\equiv \varsigma \neq 1.$

A simple two-parameter ($a$ and $\varsigma $) calculation yields the
following results\cite{clprl}:
\[
\begin{tabular}{cccc}
{\em Item} & $\chi QM\,$ & $a=0.1,\;\varsigma =-1.2$ & $\;Experimental\;value
$ \\ \hline\hline
&  &  &  \\
$I_G$ & $\frac 13-\frac 49(1-\varsigma )a$ & $0.236$ & $0.235\pm 0.026$ \\
&  &  &  \\
$\overline{u}/\overline{d}$ & $\frac{6+2\varsigma +\varsigma ^2}{8+\varsigma
^2}$ & $0.53$ & $0.51\pm 0.09$ \\
&  &  &  \\
$f_s$ & $\frac{\left( 10-2\varsigma +\varsigma ^2\right) a}{\left(
9/2\right) +3\left( 8+\varsigma ^2\right) a}$ & $0.19$ & $0.18\pm 0.03$ \\
&  &  &  \\
$\Delta u$ & $\;\;\frac 43-\frac 19\left( 37+8\varsigma ^2\right) a$ & $0.79$
& $0.83\pm 0.05$ \\
&  &  &  \\
$\Delta d$ & $-\frac 13-\frac 29\left( 1-\varsigma ^2\right) a$ & $-0.32$ & $%
-0.42\pm 0.05$ \\
&  &  &  \\
$\Delta s$ & $-a$ & $-0.10$ & $-0.10\pm 0.05$ \\
&  &  &  \\
$\Delta \Sigma $ & $1-\frac 23\left( 8+\varsigma ^2\right) a$ & $0.37$ & $%
0.31\pm 0.11$%
\end{tabular}
\]
Here $I_G$ stands for the Gottfried integral of Eq.(\ref{gsr}), and $f_s$ is
the strange quark fraction in the proton as defined in Eq.(\ref{sfrac}). The
''experimental values'' for $\Delta q\,s$ are taken from the
phenomenological analysis as given in Ref \cite{ellisk}.

Our choice of the Goldstone-mode emission-probability $a=1.0$ and the
coupling ratio $\varsigma =-1.2$ is for illustrative purpose only, to show
that a simple calculation in this model can reproduce the principal features
of the observed proton spin and flavor contents. (Given the crudity of the
model, a detailed ''best-fit'' analysis is not warranted at this stage.) It
is gratifying that the parameter $a$ turns out to be fairly small, This
means that the constituent quarks are not surrounded by some complicated
quark-sea; and it makes this chiral quark picture more self-consistent. The
negative value for the singlet-coupling $g_0\simeq -1.2g_8$, which is
basically required by the $\overline{u}/\overline{d}$ ratio as determined by
the NA51 measurement, is, in our opinion, the more intriguing result of this
model calculation.

This work is supported at CMU by the US Department of Energy
(DE-FG02-91ER-40682), and at UM-St Louis by the National Science Foundation
(PHY-9207026).

\

\end{document}